\documentclass[12pt]{iopart}

\usepackage{graphicx}

\begin{document}

\title{Entropy production and the arrow of time.}

\author{J.M.R. Parrondo$^1$, C. van den Broeck$^2$, and R. Kawai$^3$}
\address{$^1$Departamento de F{\'i}sica At{\'o}mica, Molecular
y Nuclear and GISC, Universidad Complutense de Madrid, 28040-Madrid,
Spain\\$^2$ University of Hasselt, B-3590 Diepenbeek, Belgium\\$^3$
Department of Physics, University of Alabama at Birmingham,
Birmingham, AL 35294, USA}

\ead{parrondo@fis.ucm.es}

\begin{abstract}
We present an exact relationship between  the entropy production and
the distinguishability of a process from its time-reverse,
quantified by the relative entropy between forward and backward
states. The relationship is shown to remain valid for a wide family
of initial conditions, such as canonical, constrained canonical,
multi-canonical and grand canonical distributions, as well as both
for classical and quantum systems.
\end{abstract}


\maketitle

\section{Introduction}
\label{sec:intro}

 Providing a microscopic expression  for the
entropy production has been one of the grand aims of statistical
mechanics, going back to the seminal work of Boltzmann. However,
both the range of validity of the second law and of its proposed
derivations have, from the very beginning, generated discussion and
controversy. The recent discovery of the fluctuation and work
theorems~\cite{bochkov,evans93,gallavotti95,jarzynski97,crooks,CRAS}
has re-invigorated the debate. These intriguing results provide
equalities involving the probability distributions for work or
entropy production, and appear to be of specific interest for the
study of small systems~\cite{ritort03,liphardt02,collin05}. With
respect to the average value of work or entropy production, they are
consistent with the second law, but provide no additional
information beyond it. In a recent work~\cite{kawai07} (see also
\cite {Jarzynski2006}) we have derived, from first principles, the
exact value of the average dissipated work $\langle W\rangle_{\rm
diss}$ upon bringing a system from one canonical equilibrium state
at a temperature $T$ into another one at the same temperature. The
dissipated work is defined as the extra amount of work, on top of
the difference of free energy $\Delta F$, that is required for
making this transition. Its expression reads:
\begin{equation}\label{main_formula1}
\langle W\rangle_{\rm diss}=\langle W\rangle-\Delta F= k T \left \langle \ln
\frac{\rho}{\tilde{\rho}} \right\rangle.
\end{equation}
Here $k$ is the Boltzmann constant  and $\rho=\rho(\Gamma;t)$ is the
probability density in phase space to observe the system to be in a
micro-state $\Gamma=(q,p)$ specified by a set of positions $q$ and
momenta $p$ at an intermediate time $t$. The averaging brackets
denote an average with respect to the density $\rho$.  The other
density $\tilde{\rho}=\tilde{\rho}(\tilde{\Gamma};t)$ represents the
distribution in the time-reversed process observed at a
corresponding time-reversed phase point $\tilde{\Gamma}=(q,-p)$ at
the same time as the forward process. As an example, consider
compression and expansion of a gas in a cylinder, a forward
experiment corresponding to the expansion of a gas by moving a
piston from an initial to a final position; the backward experiment
corresponds to performing the motion of the piston in the
time-reversed manner.  The statistics of the micro-states are taken
at the same intermediate position of the piston. Both forward and
backward experiment are assumed to start in canonical equilibrium at
the same temperature.

The dissipated work~(\ref{main_formula1}) can also be
written in terms of a well known and powerful concept from
information theory, the relative entropy or Kullback-Leibler
distance between two probability densities~\cite{cover-thomas}:
\begin{equation}\label{main_formula2}
\langle W\rangle_{\rm diss}= k T D(\rho\|\tilde{\rho}),
\end{equation}
where
\begin{eqnarray}
D(\rho\|\tilde{\rho})
&=& \int d\Gamma\;\rho(\Gamma,t) \ln
\frac{\rho(\Gamma,t)}{\tilde{\rho}(\tilde{\Gamma},t)}
\\
&=&\int d\Gamma \rho(\Gamma,t)\ln \rho(\Gamma,t)
-\int d\Gamma
\rho(\Gamma,t)\ln \tilde{\rho}(\tilde{\Gamma},t)\, .
\label{RE2}
\end{eqnarray}
is the relative entropy between $\rho$ and $\tilde\rho$. Written in
this way, the result reveals its deep meaning, linking the
dissipation directly to the irreversibility of the underlying
experiment. Indeed the relative entropy between probability
densities is expressing the difficulty of distinguishing samplings
from these densities~\cite{cover-thomas}. In the present case, it
measures the difficulty of distinguishing whether observed data of
the micro-state correspond to those from a forward or backward
experiment. Therefore, the relative entropy (\ref{RE2}) can be
considered as a quantitative measure of the arrow of time
\cite{Schnakenberg,Pomeau,Luo,Mackey,Qian,Jiang,Maes,gaspard04,Costa,Seifert2005,%
Porporato,Gomez,Blythe}.

At the end of the forward process, the system is generally speaking
in a nonequilibrium state, for which the entropy is not well
defined. Hence, to make the connection with the entropy production,
we introduce an idealized heat bath at temperature $T$, with which
the system is put in contact at the end of the forward experiment,
without extra work. The system thus relaxes back to canonical
equilibrium, which is also the starting state of the backward
process. During the relaxation, the system transfers an average
amount  of heat $Q$ to the bath. We can now evaluate the total
entropy production. On the one hand, the  entropy change in the
bath, assuming it operates quasi-statically,  is given by $ Q/T$. On
the other hand, the entropy change between the canonical equilibrium
state of initial and final states in the system reads: $(\Delta U-
\Delta F)/T= (\langle W\rangle-Q- \Delta F)/T$. Here we used the
relation $F=U-TS$, where $U=\langle H\rangle$ is the equilibrium
internal energy, calculated with respect to the prevailing canonical
distribution, and $S$ the thermodynamic entropy of the system. The
total entropy production (in the total device, system plus heat
bath) in the forward process, $\Delta S$, is thus equal to the
dissipated $\langle W\rangle_{\rm diss}$ divided by temperature,
hence:
\begin{equation}\label{entropy}
\Delta S=k  \left \langle \ln
\frac{\rho}{\tilde{\rho}} \right\rangle = k D(\rho\|\tilde{\rho}).
\end{equation}
This result goes right to the core of the
second law, as it provides an exact explicit  microscopic expression for the
total entropy production in the considered scenario.

This scenario is very specific: the system is initially at canonical
equilibrium, and disconnected from any heat bath during the
perturbation~\cite{kawai07}. These assumptions greatly simplify the
derivation, but also generate the misleading impression that they
are essential. The aim of this paper is to further clarify the
status of the basic results (\ref{entropy}), and to show that it has
a wider range of validity.

Our discussion will be based on a general and exact mathematical
identity for the relative entropy, derived in
Sec.~\ref{sec:general}, which  is  valid for arbitrary initial
conditions in the forward and backward process. In
Sec.~\ref{sec:scenarios}, this identity is applied  to different
scenarios, including various types of initial equilibrium states,
such as grand canonical ensembles, constrained equilibrium states,
and multi-canonical distributions at different temperatures. In each
of these cases, the entropy production is  (aside a factor $k$)
equal to the relative entropy of the phase space density between the
forward and the backward states, provided some idealized relaxation
process is assumed at the end of the forward process. In
Sec.~\ref{sec:quantum}, we present the quantum analogue of the above
result, expressed in terms of the quantum relative
entropy~\cite{vedralrep,vedral,petz}.

\section{General Formulation}
\label{sec:general}

We consider a system described by the Hamiltonian
$H(\Gamma;\lambda)$, with $\lambda$ a control parameter that
describes the energy exchange with an external device. For
simplicity, we will assume that the Hamiltonian is an even function
with respect to inversion of momenta. As is well known, the
probability density in phase space $\rho(\Gamma,t)$ to observe the
system in a specific micro-state $\Gamma$ at time $t$ obeys the
Liouville equation:
\begin{equation}
\frac{\partial\rho(\Gamma,t)}{\partial t} =
\mathcal{L}\rho(\Gamma,t)\, .
\label{L_fwd}
\end{equation}
In a classical system, the Liouville operator $\mathcal{L}$  is
given by Poisson brackets: $\mathcal{L}{\rho}=\{H,\rho\}=\partial_q
H \partial_p \rho -\partial_p H \partial_q \rho$ (summation
convention over all positions and momenta is assumed).

An important implication of the Liouville theorem is that  the
Shannon entropy associated to $\rho$, namely:
\begin{equation}\label{Gibbs}
  - \int d\Gamma \rho(\Gamma) \ln \rho(\Gamma)
\end{equation}
is constant in time, as can be easily proven by applying the
Liouville equation followed by partial integration~\cite{Mackey}.
This appears to be in contradiction with the second law of
thermodynamics, or at least precludes the use of this expression as
a microscopic definition for the entropy. In fact, we share the
general opinion that entropy is not defined at the microscopic level
except when the system is at equilibrium.  The Shannon
entropy~(\ref{Gibbs}) reduces to the proper thermodynamic entropy
only for systems at equilibrium.

An essential remark for the further discussion is that the above
property remains valid  for a smooth time-dependent Hamiltonian or
Liouvillian. We will henceforth focus on scenarios where the control
parameter $\lambda$  is changed from an initial value $\lambda_A$ to
a final value $\lambda_B$, according to a specific protocol
$\lambda(t), t \in [0,\tau]$. We also consider a time-reversed
experiment, starting at the final value $\lambda_B$ of the control
parameter, and performing the time-reversed perturbation. The
quantities appearing in this setting will be indicated by a
superscript tilde. We will always use the forward time $t$ to
designate time in both forward and backward experiment, keeping in
mind that the corresponding real time in the backward experiment is
equal to $\tau-t$. The initial conditions for both forward and
backward experiments are for the moment left unspecified. The above
set-up is reminiscent of the one  introduced by Jarzynski in his
derivation of the Jarzynski equality~\cite{jarzynski97}, although
the scope of our derivation extends, as we will see, well beyond the
validity of the Jarzynski equality.

The key observation is that the phase space density of backward evolution
with reversed momenta, namely $\tilde \rho(\tilde{\Gamma},t)$, obeys the
same Liouville equation (\ref{L_fwd}) with respect to the forward
time variable $t$:
\begin{equation}\label{L_bwd}
\frac{\partial\tilde\rho(\tilde{\Gamma},t)}{\partial t} =
\mathcal{L}\tilde\rho(\tilde{\Gamma},t)\, .
\end{equation}

The transition from
Eq.~(\ref{L_fwd}) to Eq.~(\ref{L_bwd}) is based on
micro-reversibility: trajectories retrace their steps upon reversing
the schedule of $\lambda$ and inverting momenta. Mathematically, the result
follows easily by writing the Liouville equation for $\tilde{\rho}$
in its proper time $\tau-t$. When switching to the forward time $t$,
the negative sign, due to the time derivative in the l.h.s. of the
Liouville equation, is canceled by another one in the r.h.s.,
appearing upon inverting the momenta. As a particular example, for the case of a
time-independent Hamiltonian, the above statement reduces to the fact that
$\rho({\Gamma},t)$ and $\rho({\tilde\Gamma},\tau-t)$, obey the same Liouville
equation, when the time-derivative is with respect to $t$.

We can now invoke a remarkable property for probability distributions that obey
the same Liouville equation: their relative entropy is invariant in time
\cite{Mackey,Plastino}. In particular, the following quantity:
\begin{equation} \label{crossterm2}
 \int d\Gamma \rho(\Gamma,t) \ln \tilde{\rho}(\tilde{\Gamma},t)
\end{equation}
is time-invariant, as can again be easily verified by applying Liouville's
equation followed by partial integration.

The invariance in time of (\ref{Gibbs}) and (\ref{crossterm2}) now
allows us to rewrite the quantity of central interest, namely the
relative entropy~(\ref{RE2}), as follows. Since each term in the
r.h.s. of Eq.~(\ref{RE2}) is invariant under time translation, we
can shift time in the first term back to $t=0$, and in the second
term forward to $t=\tau$:

\begin{equation}\label{general}
D \left (\rho(\Gamma,t)\| \tilde\rho(\tilde{\Gamma},t) \right ) =
 \int d\Gamma \rho(\Gamma,0)\ln \rho(\Gamma,0)
-\int d\Gamma \rho(\Gamma,\tau)\ln \tilde\rho(\tilde{\Gamma},\tau)\, .
\end{equation}
We stress again that this is an exact result, valid for any initial
conditions for $\rho$ and  $\tilde\rho$. For specific choices of
these states, we proceed to make the connection with the entropy
production in a variety of scenarios. Note also that the above
result will only be usefull if the support of $\tilde\rho$ includes
the support of $\rho$, i.e., there are no phase points for which
$\tilde\rho = 0$ and $\rho \neq  0$. Otherwise  $D (\rho\|
\tilde\rho )$ diverges.

\section{Explicit entropy production in different scenarios}
\label{sec:scenarios}

Along this section, we apply  Eq.~(\ref{general}) for specific
choices of equilibrium initial conditions $\rho(\Gamma,0)$ and
$\tilde\rho(\tilde{\Gamma},\tau)$ in forward and backward process,
respectively. We will see that in all the considered cases, the
resulting relative entropy is equal to the entropy production along
the forward process. If $\rho(\Gamma,0)$ is an equilibrium
distribution or the product of independent equilibrium
distributions, Shannon entropy is equal to the thermodynamic
entropy:
\begin{equation}
S(0)=-k\int d\Gamma\rho(\Gamma,0)\ln\rho(\Gamma,0).
\label{s0}
\end{equation}
This equality holds for different equilibrium ensembles, such as
canonical and grand canonical. With this choice of initial condition
for the forward process the first term in Eq.~(\ref{general}) is
minus the initial equilibrium entropy. In the following, we will
prove that the second term can be identified with the final entropy.

 As mentioned in
previous sections, in this paper we only consider as meaningful the
entropy of equilibrium states. On the other hand, after the forward
process is completed at time $\tau$, the resulting state
$\rho(\Gamma,\tau)$ is not at equilibrium. In order to have a well
defined expression for the entropy production, we need some
relaxation mechanism to equilibrium {\em after} the forward process
has been completed. The function of this mechanism is twofold: it
allows us to have a meaningful definition of entropy production, and
also drives the system from the non equilibrium final state
$\rho(\Gamma,\tau)$ to the equilibrium one
$\tilde\rho(\Gamma,\tau)$, which is the initial condition for the
backward process.\footnote{This relaxation will take some time
$\tau_{\rm rel}$. However, recall that our time variable $t$ in
$\tilde\rho(\Gamma,t)$ denotes the stage of the process, using the
forward one as reference, rather than the real physical time $t_{\rm
phys}$. In a complete forward-backward cycle, the time variable in
$\tilde\rho(\Gamma,t)$ will be $t=2\tau+\tau_{\rm rel}-t_{\rm
phys}$, so $\tilde\rho(\Gamma,\tau)$ is the state of the system at
$t_{\rm phys}=\tau+\tau_{\rm rel}$).}

Notice that the relaxation from $\rho(\Gamma,\tau)$ to
$\tilde\rho(\Gamma,\tau)$ in an isolated system is incompatible with
the Liouville theorem. Our derivation does not touch on this old and
unresolved problem. Instead, for the relaxation to
$\tilde\rho(\Gamma,\tau)$, we invoke the presence of one or several
ideal (super)baths, to which the system is  weakly coupled. This
relaxation can,  in principle, be performed without any extra energy
input, since coupling the superbath with just one degree of freedom
of the system is enough to induce thermalization \cite{feynman}.
Moreover, if this degree of freedom is chosen to have zero mean (for
instance, one of the momenta of the bath), then the average work
done by switching on the coupling is strictly zero. Finally, we will
assume that the entropy production in the superbaths, due to
exchange of energy and/or particles with the system, is given by the
standard expressions from statistical mechanics and thermodynamics.
Note that such or similar assumptions also appear in the various
derivations of both fluctuation and work theorems, and in the study
of nonequilibrium steady states. They are also intrinsically present
in the formulations via mesoscopic stochastic descriptions.

\subsection{Canonical distributions}

In our first example, we  start the forward process with a canonical
distribution at temperature $T$ and, after the process is completed
at time $\tau$, we connect the system with a thermal bath at a
different temperature $T'$. The initial condition for the backward
process is in this case a canonical distribution at temperature
$T'$:
\begin{eqnarray}
\rho(\Gamma,0) &=&
\frac{e^{-\beta H(\Gamma;\lambda_A)}}{Z(T,\lambda_A)} \\
\tilde\rho(\Gamma,\tau) &=&
\frac{e^{-\beta' H(\Gamma;\lambda_B)}}{Z(T',\lambda_B)}
\end{eqnarray}
$Z(T,\lambda)=\int d\Gamma e^{-\beta H(\Gamma,\lambda)}$ being the
partition function. Recalling that we consider Hamiltonians even
functions of the momenta, we obtain from Eq.~(\ref{general}) and
(\ref{s0}):
\begin{equation}
kD(\rho||\tilde\rho)=-S(0)+
\frac{\langle H\rangle_\tau-F(T',\lambda_B)}{T'}
\end{equation}
where averages $\langle \cdot \rangle_t$ are taken over the forward
time density $\rho(\Gamma,\tau)$ and
\begin{equation}
F(T,\lambda)=-kT \ln Z(T,\lambda).
\end{equation}
is the usual  equilibrium free energy. After relaxation to
equilibrium with the bath at temperature $T'$, the average energy of
the system becomes $\langle H\rangle_{{\rm eq},\tau}$, which is only
a function of $\lambda_B$ and $T'$. In the relaxation, the system
transfers an amount of energy to the thermal bath $Q=\langle
H\rangle_\tau-\langle H\rangle_{{\rm eq},\tau}$. Therefore:
\begin{eqnarray}
kD(\rho||\tilde\rho) &=& -S(0)+
\frac{\langle H\rangle_{{\rm eq},\tau}-F(T',\lambda_B)}{T'}+\frac{Q}{T'} \nonumber\\
&=& -S(0)+S(\tau)+\Delta S_{\rm bath}=\Delta S
\end{eqnarray}
where $S(\tau)$ is the final entropy of the system (in equilibrium
at temperature $T'$), and $\Delta S_{\rm bath}$ is the increase of
the entropy in the bath.

Therefore, the relative entropy is equal to the total entropy
production along the forward process plus the final relaxation:
$\Delta S =kD(\rho||\tilde\rho)$. Notice that {\em all} entropies
have been
 calculated only for equilibrium states. In fact,
this example illustrates that it makes no sense to talk about the
entropy produced along the process if one does not specify the final
equilibrium state and how is reached. In our case, the total entropy
production depends on the final temperature $T'$. Hence, the same
forward process $\lambda(t)$ can have different entropy production
depending on the final temperature $T'$.

\subsection{General equilibrium distribution}

The above example can be easily generalized to an arbitrary
equilibrium distribution defined over some conserved quantities
$A_i(\Gamma)$ ($i=1,\dots,r$), with conjugated variables
$\alpha_i=T{\partial S}/{\partial A_i}$ and the corresponding
thermodynamic potential ${\cal U}(T,\vec{\alpha})=\langle
H\rangle+\sum_i\alpha_i\langle A_i\rangle-TS$. After the forward
process, the system is coupled to a reservoir characterized by
temperature $T'$ and  certain values $\alpha'_i$ of the conjugated
variables. The initial condition for the backward process reads:
\begin{equation}
\tilde\rho(\Gamma,\tau)=\exp\left [-\beta'\left(H(\Gamma,\lambda_B)
+\sum_{i=1}^r\alpha'_i A_i(\Gamma)-{\cal U}(T',\vec{\alpha}')\right)\right]
\end{equation}
Hence
\begin{equation}
-k\int d\Gamma \rho(\Gamma,\tau)\ln
\tilde\rho(\Gamma,\tau)= \frac{ \langle H\rangle_\tau+\sum_{i=1}^r \alpha'_i
\langle A_i\rangle_\tau-{\cal U}(T',\vec{\alpha}')}{T'}
\label{cross}
\end{equation}
 During relaxation, there is
a transfer of energy and of the conserved quantities $A_i$ from the
system to the reservoir given, respectively, by $Q=\langle
H\rangle_\tau-\langle H\rangle_{{\rm eq},\tau}$ and $\Delta
A_i=\langle A_i\rangle_\tau-\langle A_i\rangle_{{\rm eq},\tau}$. The
final entropy of the system is
\begin{equation} S(\tau)=\frac{\langle H\rangle_{{\rm eq},\tau}+\sum_{i=1}^r \alpha'_i \langle A_i(\Gamma)\rangle_{{\rm eq},\tau}-{\cal
U}(T',\vec{\alpha}')}{T'}
\end{equation}
and the entropy increase in the reservoir is $\Delta S_{\rm
bath}=(Q+\sum_i\alpha'_i\Delta A_i)/T'$. The cross term
(\ref{cross}) is then equal to $S(\tau)+\Delta S_{\rm bath}$. If,
along the forward process, the external agent moving $\lambda (t)$
does not undergo any change of entropy,  we again find
Eq.~(\ref{entropy}): $\Delta S=kD(\rho||\tilde \rho)$.

Notice that the cross entropy (\ref{cross}) contains precisely the
non equilibrium average values, $\langle H\rangle_\tau$ and $\langle
A_i(\Gamma)\rangle_\tau$, yielding the correct expression for the
final equilibrium entropy of the system $S(\tau)$ {\em plus} the
entropy increase in the bath $\Delta S_{\rm bath}$.

As a particular case, consider  $A=N$ the number of particles. Then,
$-\alpha=\mu$,  the chemical potential. In equilibrium, the
thermodynamic potential
\begin{equation}
 {\cal U}(T,\mu)=\langle H\rangle_{\rm eq}-\mu\langle
 N\rangle_{\rm eq}-TS
\end{equation}
is the so-called grand canonical potential, where the averages
$\langle\cdot\rangle_{\rm eq}$ are taken with respect to the grand
canonical ensemble \begin{equation}
\rho(\Gamma,N)=e^{-\beta[H(\Gamma,\lambda)-\mu N-{\cal U}(T,\mu)]}
\end{equation}
We now choose as initial condition for the forward process a grand
canonical ensemble characterized by temperature $T$ and chemical
potential $\mu$. We proceed with the forward process by changing
$\lambda(t)$ ($N$ remaining constant to avoid change of entropy in
the external agent moving $\lambda$). After the process is completed
we connect the system with a particle and energy reservoir at
temperature $T'$ and chemical potential $\mu'$. In the relaxation,
there will be an exchange of energy and particles between the system
and the bath, as explained above. Finally, the total entropy
production along the forward process plus the relaxation is given by
the relative entropy between forward and backward states,
Eq.~(\ref{entropy}).

\subsection{Multicanonical distribution}

The above examples can be extended to the product of several
equilibrium states. We consider the special case where systems at
different temperatures are brought into contact with each other.
This case is of particular interest since thermal exchange is
another standard example of entropy production. Also, the present
scenario produces in the appropriate limit, e.g., the limit of
infinitely large thermal reservoirs, a stationary nonequilibrium
state. We will thus show that our formula of entropy production as a
relative phase space entropy also applies in this example of a
nonequilibrium steady state.

Consider $n$  systems $i=1,\cdots,n$, at the initial time decoupled,
and each one at canonical equilibrium at temperature  $T_i$:
\begin{equation}
\rho(\Gamma,0)=\prod_{i=1}^n\frac{e^{-\beta_i
H_i(\Gamma_i,\lambda_A)}}{Z_i(T_i,\lambda_A)}
\label{mcinif}
\end{equation}
We will assume a similar initial state for  the backward process,
decoupled systems but with arbitrary temperatures $T_i'$, i.e.:
\begin{equation}
\tilde\rho(\Gamma,\tau)=\prod_{i=1}^n\frac{e^{-\beta_i'
H_i(\Gamma_i,\lambda_B)}}{Z_i(T'_i,\lambda_B)}.\label{mcinib}
\end{equation}
While the systems are  decoupled both at initial and final times,
thermal contact with each other can be established during
intermediate stages of the process by an appropriate choice of the
time-dependent  Hamiltonian. One possibility is adding to the
original decoupled Hamiltonians a coupling term inducing energy
exchange and multiplied by the control parameter $\lambda(t)$, set
equal to zero at initial and final times. Eq.~(\ref{general}) gives
us the following result:
\begin{eqnarray}
k D(\rho\|\tilde\rho)&=&\sum_{i=1}^n \left[-\frac{\langle
H_i(\lambda_A)\rangle_0}{T_i}- k\ln
Z_i(T_i,\lambda_A)\right]\nonumber
\\&&+\sum_{i=1}^n \left[\frac{\langle
H_i(\lambda_{B})\rangle_{\tau}}{T'_i}+k \ln
Z_i(T',\lambda_{B})\right]
\label{multican0}
\end{eqnarray}
As in our previous examples, each subsystem transfers to their
corresponding thermal baths an energy $Q_{i}= \langle
H_i(\lambda_{B})\rangle_{\tau}-\langle H_i(\lambda_B)\rangle_{{\rm
eq},\tau}$, $\langle H_i(\lambda_B)\rangle_{{\rm eq},\tau}$ being
the equilibrium energy of each subsystem after the relaxation to
$\tilde\rho(\Gamma,\tau)$. Therefore:
\begin{eqnarray}
k D(\rho\|\tilde\rho)&=&\sum_{i=1}^n \left[-\frac{\langle
H_i(\lambda_A)\rangle_0-F(T_i,\lambda_A)}{T_i}+
\frac{\langle
H_i(\lambda_{B})\rangle_{{\rm eq},\tau}-F(T_i',\lambda_B)+Q_{i}}{T'_i}\right]\nonumber \\ &=&
\sum_{i=1}^n \left[-S_i(0)+S_i(\tau)+\Delta S_{{\rm bath},i}\right] = \Delta S
\label{multican}
\end{eqnarray}
 Hence, the
total entropy produced in the entire forward process (plus
relaxation) is again given by the relative entropy,
Eq.~(\ref{entropy}). Similar arguments can be applied to multi-grand
canonical distributions, allowing chemical reactions and exchange of
particles between the system and the final reservoirs.

\subsection{Constrained canonical distributions and filters}

Finally, we now apply the general result Eq.~(\ref{general}) to the
case of canonical states restricted to subdomains of the phase
space. One can imagine two scenarios leading to such restricted
initial states, namely: (a) the system is subject to constraints in
the initial and/or final states, or (b) one selects (or restricts
the observation to) that specific set of trajectories that lie
within a prescribed subdomain of phase space; this selection can, in
principle, be performed at any intermediate time of the process.

In both cases, the relevant initial  condition for the forward
process can be written as follows:
\begin{equation}
\rho(\Gamma,0)=
\frac{ e^{-\beta
H(\Gamma;\lambda_A)}}{Z(\beta,\lambda_A;D_A)}\chi_{D_A}(\Gamma)
\end{equation}
where the characteristic function is defined as
\begin{equation}
\chi_D(\Gamma) =\left\{
\begin{array}{cl}
1 & \Gamma \in D\\
0 & {\rm otherwise}
\end{array}\right.
\end{equation}
for any subset $D$ in the phase space.  The partition function is
then defined as $Z(\beta,\lambda;D)=\int_D d\Gamma\,e^{-\beta
H(\Gamma;\lambda)}$.

As initial condition for the backward process we consider:
\begin{equation}
\tilde\rho(\Gamma,\tau)=
\frac{e^{-\beta
H(\Gamma;\lambda_B)}}{Z(\beta,\lambda_B;{D}_B)}\chi_{{D}_B}(\Gamma)
\end{equation}
The prescriptions for $D_A$ and $ D_B$ are as follows. For case (a),
the subsets $D_A$ and $D_B$ reproduce the imposed initial
constraints for the forward and backward processes, respectively.
For case (b), the set of filtered forward trajectories equals $D_A$
at the initial time $t=0$ and $D_B$ at the final time $\tau$. For
simplicity, we will assume that $D_B$ is invariant under inversion
of momenta $p\to -p$.

The relative entropy (\ref{general}) now reads:
\begin{eqnarray}
kT\, D(\rho\|\tilde\rho) &=&
 \langle H(\Gamma;\lambda_B)\rangle_\tau
-\langle H(\Gamma;\lambda_A)\rangle_0  -kT \ln\frac{Z(\beta,\lambda_A;D_A)}{Z(\beta,\lambda_B; D_B)}
\nonumber\\
&=& \langle W \rangle
-kT\ln\frac{Z(\beta,\lambda_A;D_A)}{Z(\beta,\lambda_B; D_B)}
\end{eqnarray}
There are two possible interpretations  of this result. If the
restriction of microstates to $D_A$ and $D_B$ is due to a filter, as
in scenario (b), the initial and final free energies must be
calculated over the whole phase space. In this case, we have
\begin{equation}
kT\, D(\rho\|\tilde\rho) =\langle W\rangle - \Delta F
- kT\ln \left [ \frac{Z(\beta,\lambda_A;D_A)Z(\beta,\lambda_B)}
 {Z(\beta,\lambda_A)Z(\beta,\lambda_B;D_B)} \right ]
\end{equation}
The ratios between the restricted and the  non-restricted partition
functions are precisely the probability that an arbitrary trajectory
passes the filter in the forward ($p_{D_A}$) and backward processes
($\tilde p_{D_B}$), respectively. Therefore, we obtain the following
exact equality:
\begin{equation}
kT\, D(\rho\|\tilde\rho)+kT\ln\frac{p_{D_A}}{\tilde p_{D_B}} = \langle
W\rangle-\Delta F \,.
\end{equation}
In particular, the non-negativeness of the relative entropy yields the following
inequality:
\begin{equation}
\langle W \rangle-\Delta F \ge
kT \ln\frac{p_{D_A}}{\tilde p_{D_B}}
\end{equation}
which was derived in \cite{kawai07} using the Liouville  theorem and
applied to a Brownian version of the Szilard engine and the
restore-to-zero process introduced by Landauer.

On the other hand, if the restriction is  due to constraints in the
system at $t=0$ and $\tau$, the logarithm of the restricted
partition functions can be considered as the actual initial and
final free energies. Hence, we recover the same result as in the canonical case:
\begin{equation} \label{constrained}
kT\, D(\rho\|\tilde\rho) =\langle W \rangle - \Delta F=
\langle W \rangle_{\rm diss}
\end{equation}

It is worthwhile to illustrate the last case on a concrete example,
especially since  the Jarzynksi and Crooks equalities are failing
for this kind of scenario \cite{jarzynskifail}. We consider the case
of free expansion of a gas from a volume $V/2$ to a final volume
$V$. The Hamiltonian in this case is time-independent, but we can
still apply our theory to this process consisting of a free
relaxation from a restricted initial condition. In the forward
process, the gas is initially in canonical equilibrium, but confined
to one half of the container, with vacuum in the other half. It is
subsequently allowed to expand and fill the entire container. Since
the Hamiltonian is time independent, the backward process starts
with the gas at equilibrium filling the whole volume $V$ of the
container, and obviously remaining so for all times. We thus have
\begin{eqnarray}
\rho(\Gamma,0)
&=&\frac{e^{-\beta H}}{Z(\beta;V/2)}\chi_{V/2}(\Gamma)
\nonumber\\
\tilde\rho(\tilde\Gamma,\tau)
&=&\frac{e^{-\beta H}}{Z(\beta;V)}\chi_{V}(\Gamma)
\nonumber
\end{eqnarray}
where the characteristic function $\chi_V(\Gamma)$  is 1 for
microstates with all the particles in the volume $V$ and 0
otherwise. Now the relative entropy is
\begin{equation}
kT\, D(\rho\|\tilde\rho)
= \langle H \rangle_\tau -\langle H \rangle_0 - kT\ln
\frac{Z(\beta;V/2)}{Z(\beta;V)}
= - \Delta F
\end{equation}
since the energy is conserved along the  evolution ($H$ has no
explicit time-dependence). There is no work performed in this
experiment, $\langle W \rangle=0$, and thus $-\Delta F$ is the
dissipative work (the work that we waste compared to a reversible
expansion). Therefore, Eqs.~(\ref{main_formula1}) and
(\ref{entropy}) remain valid, even though the Jarzynski and Crooks
relations no longer hold.

\section{Quantum version}
\label{sec:quantum}

Just as the Jarzynski equality has been  shown to remain valid for
quantum
processes~\cite{Tasaki,Kurchan,yukawa00,mukamel03,chernyak04,deroeck04,monnai05,
esposito06,Engel,Talkner,andrieux2008}, we now show that  our
approach can easily be extended to the quantum. Time evolution in
quantum mechanics is ruled by unitary transformations. In particular
the density operator obeys the following quantum version of the
Liouville equation:
\begin{equation}
\frac{\partial}{\partial t}\rho=\mathcal{L}\rho=-\frac{i}{\hbar}[H,\rho]
\end{equation}
Under this unitary evolution,  the Von Neumann entropy $S_{\rm
N}(\rho)=-k{\rm Tr} ( \rho \ln \rho )$ is preserved. For the
time-reversed process, the time-reversed density operator
$\tilde\rho = \Theta\rho\Theta^{-1}$ satisfies the same quantum
Liouville equation if the forward time $t$ is used (the actual time
is $\tau-t$ in the same sense as in the classical case). Here
$\Theta$ is the anti-linear time-reversal operator and a fundamental
symmetry of the Hamiltonian $\Theta H(t) = H(t)\Theta$ is
assumed~\cite{sisterna00}. Noting that $\Theta p \Theta^{-1} = -p$,
the commutation of $H$ and $\Theta$ is, in the present context,
equivalent to the assumption that the classical Hamiltonian is
symmetric with respect to the inversion of momenta.  Then, we can
prove that ${\rm Tr} (\rho \ln \tilde\rho )$ is also time invariant.
In general, both von Neumann entropy and the quantum relative
entropy are invariant under any unitary transformation~\cite{petz}.
Hence, each term in the quantum version of the relative entropy
\cite{vedralrep}:
\begin{equation}\label{quantum}
D_{\rm Q}(\rho(t)\|\tilde\rho(t))= {\rm Tr} (\rho(t) \ln \rho(t)
)-  {\rm Tr} (\rho(t) \ln \tilde{\rho}(t) ),
\end{equation}
can again be shifted to the initial and final time, respectively:
\begin{equation}\label{general_QM}
D_{\rm Q}(\rho \| \tilde{\rho} ) = {\rm Tr} [\rho(0)\ln\rho(0) ]
-{\rm Tr}[\rho(\tau) \ln \tilde \rho(\tau) ],
\end{equation}
which is analogous to the classical result (\ref{general}).

By using the appropriate choices of initial and final equilibrium
density matrices,  one can easily derive the quantum versions of all
the aforementioned classical results.  As an illustration, we
present the quantum version of Eq.~(\ref{main_formula1}).
Substituting canonical equilibrium density operators
\begin{equation}
\rho(0) = \frac{-e^{\beta H(\lambda_A)}}{Z(T,\lambda_A)}, \quad
\tilde\rho(\tau) = \frac{-e^{\beta H(\lambda_B)}}{Z(T,\lambda_B)}
\end{equation}
where the partition function is $Z(T,\lambda)={\rm Tr}\, e^{-\beta
H(\lambda)}$.  into Eq.~(\ref{general_QM}), we  again obtains
\begin{equation}
kT D_{\rm Q}(\rho \| \tilde{\rho} )=\langle W\rangle-\Delta
F=\langle W\rangle_{\rm diss}\,.
\end{equation}


\section{Measuring the relative entropy}
\label{sec:measuring}

While our general formula for entropy production is exact and may be
useful as the starting point for other theoretical investigations,
the experimental or even numerical measurement of the phase space
density may practically not be feasible. As shown in
Ref.~\cite{kawai07}, one can still have a good estimate of the
relative entropy through the measurement of a limited number of
physical quantities. With this partial information, it is clear that
the distinction between the forward and backward processes becomes
harder.
Mathematically, this is tantamount to saying that the relative
entropy decreases when only partial information of the system is
used:
\begin{equation}
 \Delta S = k D(\rho\|\tilde{\rho}) \ge k D(p\|\tilde{p})\, .
 \label{ineq}
\end{equation}
where $p$ and $\tilde p$ are the probability distributions of the
quantities used to describe the system in the forward and backward
process, respectively. We have checked the accuracy of this
inequality in different situations \cite{kawai07,gomezpre,Gomez}. It
is remarkable that, in the canonical scenario with $T=T'$, the
variables coupled with $\lambda(t)$ are enough to saturate the
inequality \cite{Gomez}.

The quantum case is more involved. One can lose information about
the state of the system in different ways: by performing specific
measurements or by tracing out the degrees of freedom of a
subsystem, such as a thermal bath. In any of these two cases, we can
derive an inequality similar to (\ref{ineq}).

Consider first a measurement of the observable $\hat{\Omega}=\sum_i
\omega_i \mathcal{P}_i$ where the projection operator satisfies the
closure relation: $\sum_i \mathcal{P}_i=I$. The probability to
obtain $\omega_i$ is given by $p_i={\rm Tr} (\rho \mathcal{P}_i)$
and $\tilde{p}_i={\rm Tr} (\tilde\rho \mathcal{P}_i)$ for the
forward and backward processes, respectively.  Then, from the joint
convexity of the relative entropy~\cite{petz}, we have
\begin{equation}\label{coarse}
\Delta S= kD_{\rm Q}(\rho\|\tilde\rho) \ge kD(p\|\tilde{p}).
\end{equation}
where the discrete version of classical relative entropy is defined as
\begin{equation}
 D(p\|\tilde p)=\sum_i p_i \ln \frac{p_i}{\tilde{p}_i} \, .
\end{equation}

It should be stressed that the corresponding von Neumann entropy shows the
opposite inequality, namely,
\begin{equation}
 S(p) \ge S_{\rm N}(\rho)\,.
\end{equation}
Since some information is lost, the Shannon  information entropy
increases.  However, this result should not be interpreted as the
increase of thermodynamic entropy due to the quantum measurement
since the quantum states after measurement are not in thermal
equilibrium and the Shannon entropy does not correspond to
thermodynamic entropy. The inequality merely indicates that an
estimate of von Neumann entropy using the probability distribution
of measured quantities provides an upper bound.

Next we consider compound systems.   For simplicity, we consider a
bipartite system consisting of two subsystems $a$ and $b$. The total
Hilbert space is the product space $\mathcal{H}_{a+b}=\mathcal{H}_a
\otimes \mathcal{H}_b$. If we are interested in the state of the
subsystem $a$, the other subsystem can be traced out to obtain a
reduced density operator $\rho_a={\rm Tr}_b\, \rho_{a+b}$. The
monotonicity of the relative entropy states \cite{vedralrep}:
\begin{equation}
\Delta S=
kD_{\rm Q}(\rho_{a+b}\|\tilde{\rho}_{a+b}) \ge kD_{\rm Q}(\rho_a \|
\tilde{\rho}_a).
\end{equation}
Since we lose some information about the  system, again we cannot
find the exact amount of the dissipation. However, the relative
entropy of reduced densities provides its lower bound.

While the quantum version of the relative  entropy  shows the same
properties as the classical version, it is, again, worth mentioning that the
von Neumann entropy satisfies an unusual inequality:
\begin{equation}
|S_{\rm N}(\rho_a) -S_{\rm N}(\rho_b)|\,\leq \, S_{\rm N}(\rho_{a+b}) \, \leq \,
S_{\rm N}(\rho_a)+S_{\rm N}(\rho_b).
\end{equation}
This property of sub-additivity is quite  remarkable since, unlike
the classical Shannon entropy, the entropy of a subsystem can be
larger than that of the whole system.  This observation plays an
important role in quantum information theory.  Furthermore, the role
of such specific quantum features on the foundations of statistical
mechanics is the object of an ongoing
debate~\cite{germmer,popescu06,goldstein06,reimann08}.

\section{Conclusions}

We have shown that the relative entropy or Kullback-Leibler distance
between the forward and backward state is equal to the entropy
production along the forward processes in different scenarios, both
classical and quantum. Stein's lemma \cite{cover-thomas} gives a
precise meaning to the relative entropy $D(\rho_A||\rho_B)$ between
two distributions $\rho_A$ and $\rho_B$: if $n$ data from $\rho_B$
are given, the probability of guessing incorrectly that the data
come from $\rho_A$ is bounded by $2^{-nD(\rho_A||\rho_B)}$, for $n$
large. Therefore, the relative entropy is a quantitative measure of
the distinguishibility between the two distributions. In the case of
the forward $\rho(t)$ and backward $\tilde\rho(t)$ states of a
system, the relative entropy $D(\rho(t)||\tilde\rho(t))$ measures
the distinguishibility of the direction of time. Therefore, our main
result $\Delta S=kD(\rho(t)||\tilde\rho(t))$, is a quantitative
relationship between entropy production and the arrow of time.

We have proven this identity in a number of scenarios, by choosing
the appropriate initial conditions for the forward and the backward
process. The identity is valid whenever $\rho(0)$ and
$\tilde\rho(\tau)$ are given by equilibrium states or factorized
equilibrium states (such as in the multicanonical case).

From a fundamental point of view, one can of course argue, as we
already mentioned in the introduction, that we have merely shifted
the issues of relaxation and entropy production to the purported
properties of the bath. While this is, strictly speaking, indeed the
case, the resulting expression of the entropy production in terms of
relative entropy, Eq.~(\ref{entropy}), incorporates  two fundamental
properties, namely time-translational invariance (Liouville
equation) and time-reversibility  of the laws of physics. Hence, our
explicit expression for the entropy production quantified as the
statistical time asymmetry, is very much in the spirit of Onsager's
work on the symmetry of the Onsager coefficients, resulting from the
same ingredients, Liouville's theorem and micro-reversibility.


Finally, in the quantum case, while the extension of our results to
quantum processes looks straightforward, at least mathematically,
there are intriguing fundamental questions. For example, how do
decoherence, quantum measurement and quantum entanglement contribute
to the fluctuation theorems and ultimately to the second law? On the
other hand, the von Neumann entropy and the quantum relative entropy
can detect such processes through the change in the density
operators. In fact, these entropies are used to quantify the degree
of entanglement in quantum information theory~\cite{petz,vedral}.
Therefore, our novel approach for irreversibility in quantum
processes allows one to investigate quantum non-equilibrium
thermodynamics involving entanglement and decoherence from a novel
perspective.

\ack

We thank fruitful discussions with R. Brito, R. Marathe, and E.
Rold\'an. J.M.R.P. acknowledges financial support from Spanish
Ministerio de Ciencia Innovaci\'on through grant MOSAICO.


\section*{References}


\begin{thebibliography}{99}

\bibitem{bochkov}
 G. N.Bochkov and Y. E. Kuzovlev, Physica   \textbf{A106}, 443
(1981) ; 480 (1981).

\bibitem{evans93}
D.~Evans, E.~G.~D.~Cohen, and G.~P.~Morris, Phys. Rev.
Lett. \textbf{71}, 2401 (1993).

\bibitem{gallavotti95}
G.~Gallavotti and E.~G.~D.~Cohen, Phys. Rev.
Lett. \textbf{74}, 2694 (1995).

\bibitem{jarzynski97}
C.~Jarzynski, Phys. Rev. Lett. \textbf{78}, 2690 (1997);
Phys. Rev. E \textbf{56}, 5018 (1997).

\bibitem{crooks}
G.~E.~Crooks, J. Stat. Phys. \textbf{90}, 1481 (1998); Phys. Rev. E \textbf{60},
2721 (1999).

\bibitem{CRAS}
B.~Derrida, P.~Gaspard, and C.~Van den Broeck, Editors, "Work, dissipation, and
fluctuations in nonequilibrium physics", special issue {\it C.R. Physique}
\textbf{8} (2007).

\bibitem{ritort03} F.~Ritort, Semin. Poincare  \textbf{2}, 195 (2003).

\bibitem{liphardt02} J.~Liphardt, S.~Dumont, S.~B.~Smith, I.~Tinoco (Jr) and
C.~Bustamante, Science  \textbf{296}, 1832 (2002).

\bibitem{collin05}
D.~Collin, F.~Ritort, C.~Jarzynski, S.~B.~Smith, I.~Tinoco
(Jr), and C. Bustamante, Nature  \textbf{437}, 231 (2005).

\bibitem{kawai07}
R.~Kawai, J.~M.~R.~Parrondo, and C.~Van~den~Broeck, Phys. Rev. Lett.
\textbf{98}, 080602 (2007).

\bibitem{Jarzynski2006}
C.~Jarzynski, Phys. Rev. E \textbf{73}, 046105 (2006).

\bibitem{cover-thomas}
T.~M.~Cover and J.~A.~Thomas, {\em Elements of Information Theory},
(Wiley, Hoboken, New Jersey, 2nd ed., 2006).

\bibitem{Schnakenberg}
J.~Schnakenberg, Rev. Mod. Phys.  \textbf{48}, 571 (1976).

\bibitem{Pomeau}
Y.~Pomeau, J. Phys. \textbf{43}, 859 (1982).

\bibitem{Luo}
J.L.~Luo, C.~Van~den~Broeck and G.~Nicolis, Z.
Phys. \textbf{B56}  165 (1984).

\bibitem{Mackey}
M.~C.~Mackey, Rev. Mod. Phys.  \textbf{61}, 981 (1989);
{\em Time's Arrow: The Origins of Thermodynamic Behavior},
(Springer-Verlag, New York, 1992).

\bibitem{Qian}
H.~Qian, Phys. Rev. E \textbf{63},  042103 (2001).

\bibitem{Jiang}
D.Q.~Jiang, M.~Qian, and F.-X.~Zhang, J. Math. Phys.
\textbf{44}, 4176 (2003).

\bibitem{Maes}
C.~Maes and K.~Netoc�yny, J. Stat. Phys. \textbf{110},
269 (2003).

\bibitem{gaspard04}
P.~Gaspard, J. Stat. Phys. \textbf{117}, 599 (2004).

\bibitem{Costa}
M.~Costa, A.~Goldberg, and C.-K.~Peng, Phys. Rev. Lett.
\textbf{95}, 198102 (2005).

\bibitem{Seifert2005}
U.~Seifert, Phys. Rev. Lett. \textbf{95},  040602 (2005).

\bibitem{Porporato}
A.~Porporato, J.~R.~Rigby, and E.~Daly
Phys. Rev. Lett.
\textbf{98}, 094101 (2007).

\bibitem{Gomez}
A.~Gomez-Marin, J.~M.~R.~Parrondo and C.~Van~den~Broeck, EPL \textbf{82}, 50002
(2008).

 \bibitem{Blythe}
 R.~A.~Blythe, Phys. Rev. Lett. \textbf{100},  010601 (2008).

\bibitem{vedralrep} V.~Vedral, Rev. Mod. Phys. {\bf 74}, 197 (2002).


\bibitem{vedral}
V.~Vedral, {\em Introduction to Quantum Information Science},
(Oxford Univ. Press, Oxford, 2006).

\bibitem{petz}
D. Petz, {\em Quantum Information Theory and Quantum Statistics},
(Springer, Berlin Heidelberg, 2008).



\bibitem{Plastino}
A.~R.~Plastino adn A.~Daffershofer, Phys. Rev. Lett. \textbf{93},  138701
(2004).

\bibitem{feynman} J.M.R. Parrondo and P. Espa\~{n}ol, Am. J. Phys. {\bf 64}, 1125 (1996).

\bibitem{jarzynskifail}
D.H.E. Gross, cond-mat/0508721 (2005),  C. Jarzynski,
cond-mat/0509344 (2005);  see also:
 I. Bena, C. Van den Broeck and R. Kawai,
EPL  \textbf{71}, 879 (2005).

\bibitem{Tasaki}
H.Tasaki, cond-mat/0009244 (2000).

\bibitem{Kurchan}
J. Kurchan, cond-mat/0007360 (2000).

\bibitem{yukawa00}
S. Yukawai, Jpn. J. Phys. Soc. \textbf{69}, 2367 (2000).

\bibitem{mukamel03}
S. Mukamel, Phys. Rev. Lett. \textbf{90}, 170604 (2003).

\bibitem{chernyak04}
V. Chernyak and S. Mukamel, Phys. Rev. Lett. \textbf{93}, 048302 (2004).

\bibitem{deroeck04}
W. De Roeck and C. Maes, Phys. Rev. E \textbf{69}, 026115 (2004).

\bibitem{monnai05}
T. Monnai, Phys. Rev.  E \textbf{72}, 027102 (2005).

\bibitem{esposito06}
M. Esposito and S. Mukamel, Phys. Rev. E \textbf{73}, 046129 (2006).

\bibitem{Engel}
A. Engel and R. Nolte, EPL  \textbf{79}, 10003 (2007).

\bibitem{Talkner}
P. Talkner, E. Lutz, and P. Hanggi,  Phys. Rev. E \textbf{75}, 050102 (2007).

\bibitem{andrieux2008}
D. Andrieux and P. Gaspard, Phys. Rev. Lett. \textbf{100}, 230404 (2008).

\bibitem{gomezpre}
A Gomez-Marin, J.M.R. Parrondo and C. Van den Broeck, Phys. Rev. E \textbf{78},
011107 (2008).

\bibitem{jarzynski2009}
J. Horowitz and C. Jarzynski,  cond-mat/0901.0576v1

\bibitem{sisterna00}
P. D. Sisterna, Fund. Phys. Lett. \textbf{13}, 205 (2000).

\bibitem{germmer}
J. Germmer, M. Michel, and G. Mahler, {\em Quantum Thermodynamics},
(Springer, Berlin Heidelberg, 2004).

\bibitem{popescu06}
S. Popescu, A. J. Short, and A. Winter, Nature Physics \textbf{2}, 758
(2006).

\bibitem{goldstein06}
S. Goldstein, J. L. Lebowitz, R. Tumulka, and N. Zanghi, Phys. Rev.
Lett. \textbf{96}, 050403 (2006).

\bibitem{reimann08}
P. Reimann,  Phys. Rev.
Lett. \textbf{101}, 190403 (2008).

\end{thebibliography}
\end{document}